\documentclass[printer]{aa} 

\usepackage{graphicx}
\usepackage{txfonts}
\usepackage{xcolor}

\usepackage{soul}

\newcommand\bNabla{\boldsymbol{\nabla}}

\long\def\comment#1{}

\begin{document}

\title{Helios 2 observations of solar wind turbulence decay in the inner heliosphere}
\titlerunning{Turbulence decay in the inner heliosphere}

\author{L. Sorriso-Valvo\inst{1,2,3}\fnmsep\thanks{Corresponding author: lucasorriso@gmail.com}, R. Marino\inst{4}, R. Foldes\inst{4,5}, E. L\'ev\^eque\inst{4}, R. D'Amicis\inst{6}, R. Bruno\inst{6}, D. Telloni\inst{7} \and E. Yordanova\inst{1}}

\institute{Swedish Institute of Space Physics (IRF), \r{A}ngstr\"om Laboratory, L\"agerhyddsv\"agen 1, 75121 Uppsala, Sweden
\and
CNR, Istituto per la Scienza e la Tecnologia dei Plasmi, via Amendola 122/D, 70126, Bari, Italy
\and
Space and Plasma Physics, School of Electrical Engineering and Computer Science, KTH Royal Institute of Technology, Teknikringen 31, SE-11428 Stockholm, Sweden
\and
Universit\'e de Lyon, CNRS, \'Ecole Centrale de Lyon, INSA Lyon, Universit\'e Claude Bernard Lyon 1, Laboratoire de M\'ecanique des Fluides et d'Acoustique,  UMR5509, F-69134 \'Ecully, France
\and
Dipartimento di Scienze Fisiche e Chimiche, Universit\`a degli Studi dell'Aquila, Via Vetoio 42, I-67100 Coppito AQ, Italy
\and
National Institute for Astrophysics (INAF) -– Institute for Space Astrophysics and Planetology (IAPS), Via Fosso del Cavaliere, 100, 00133 Rome, Italy
\and
National Institute for Astrophysics (INAF) -- Astrophysical Observatory of Torino, Via Osservatorio 20, I-10025 Pino Torinese, Italy
}
\authorrunning{Sorriso-Valvo et al.}
\date{Received September 15, 1996; accepted March 16, XXX}

 
  \abstract
   {}
   {The linear scaling of the mixed third-order moment of the magnetohydrodynamic fluctuations is used to estimate the energy transfer rate of the turbulent cascade in the expanding solar wind.}
   {In 1976 the Helios 2 spacecraft measured three samples of fast solar wind originating from the same coronal hole, at different distance from the sun. Along with the adjacent slow solar wind streams, these represent a unique database for studying the radial evolution of turbulence in samples of undisturbed solar wind. A set of direct numerical simulations of the MHD equations performed with the Lattice-Boltzmann code FLAME is also used for interpretation.}
   {We show that the turbulence energy transfer rate decays approximately as a power law of the distance, and that both the amplitude and decay law correspond to the observed radial temperature profile in the fast wind case. Results from magnetohydrodynamic numerical simulations {of decaying magnetohydrodynamic turbulence show a similar trend for the total dissipation, suggesting an interpretation of the observed dynamics in terms of decaying turbulence, and that multi-spacecraft studies of the solar wind radial evolution may help clarifying the nature of the evolution of the turbulent fluctuations in the ecliptic solar wind.}}
   {}


   \maketitle
%

\section{Introduction}


%
%
Spacecraft observations of interplanetary fields and plasma show that the solar wind is highly turbulent \citep{BrunoCarbone2013}. 
After the onset of the turbulent cascade at coronal level \citep{Kasper2021,Bandyopadhyay2022,Zhao2022}, several processes may energize the fluctuations during the solar-wind expansion \citep{Verscharen2019}: nonlinear decay of {large-scale} Alfv\'en waves { of solar or coronal origin} \citep{Malara2000,Chandran2018}, expansion-related and coronal-driven shears \citep{Velli1990,Tenerani2017}, pick-up ions interaction, magnetic switchbacks \citep{Bale2021,Hernandez2021,Telloni2022SB}, large-scale structures and instabilities \citep{Roberts1992,Kieokaew2021}.
The properties of turbulence are strongly variable \citep{BrunoCarbone2013}, reflecting the diversity of solar coronal sources, which modulate density, velocity, temperature and ion composition of the plasma \citep{VonSteiger2000}. 
{ Solar wind intervals are often classified according to their bulk speed, $V_{sw}$, as fast ($V_{sw}\gtrsim$ 600 km s$^{-1}$) or slow ($V_{sw}\lesssim$ 500 km s$^{-1}$).
However, turbulence properties more clearly depend on the Alfv\'enic nature of the fluctuations, for example measured using the normalized cross-helicity, $\sigma_c=\langle \delta \mathbf{v} \cdot \delta \mathbf{b} \rangle/\langle |\delta \mathbf{v}|^2+ |\delta \mathbf{b}|^2 \rangle$, where the magnetic field $\mathbf{B}$ is transformed in velocity units through the mass density $\rho$, $\mathbf{b}=\mathbf{B}/(4\pi\rho)^{1/2}$ \citep{Matthaeus1982b}, $\delta$ indicates fluctuations with respect to the mean, and brackets indicate sample average. 
For example, large-scale Alfv\'enic fluctuations may reduce the nonlinear energy transfer by sweeping apart the interacting structures \citep{Kraichnan1965,Dobrowolny1980}. }

%
%
In non-Alfv\'enic solar wind { (often observed in slow intervals)}, the turbulence generates broadband power-law magnetic spectra, $E(f)\sim f^{-\alpha}$.  
Scaling exponents $\alpha$, close to Kolmogorov's 5/3 \citep{Kolmogorov1941}, are observed from the injection scales of solar wind structure ($\sim$hours), to the characteristic ion scales ($\sim$10 s at 1 au)~\citep[][]{Bruno2014b}, where field-particle effects become relevant and the spectral exponents increase~\citep[][]{Leamon1998}.
Within such broad inertial range, strong intermittency is observed \citep{Sorriso-Valvo1999}, revealing inhomogeneously distributed small-scale structures, such as vortices and current sheets, generated by the nonlinear interactions \citep{Salem2009,Greco2016}. 
These characteristics are robustly observed at any distance from the Sun beyond 0.3 au, with no or limited radial evolution of spectral range extension or intermittency \citep{Bruno2003}.

%
%
{ On the other hand, in typically fast Alfv\'enic wind strongly aligned low-frequency velocity and magnetic fluctuations produce a $1/f$ magnetic spectral range \citep{Matthaeus&Goldstein1986,Verdini2012,Chandran2018,Matteini2018}. 
The classical turbulence inertial range is narrower, with spectral exponents between 5/3 and the Kraichnan's 3/2 \citep{Kraichnan1965}.
Note that Alfv\'enic slow solar wind was recently abundantly observed close to the Sun \citep{Chen2020}, and less frequently near 1 au \citep{DAmicis2021b}. 
In the Alfv\'enic wind, turbulence shows a clearer evolution as the wind radially expands, with
the $1/f$ break drifting towards lower frequencies \citep{bavassano1982a}. 
The inertial range broadening is interpreted as the growing Reynolds' number \citep{Frisch1995,Parashar2019}, which together with the observed increasing intermittency \citep{Bruno2003,Bruno2014}, would suggest an evolution of the solar wind dynamics towards more developed turbulence states. \citep{Tu1995,Bavassano2002,Burlaga2004,Macek2012,BrunoCarbone2013,Fraternale2016,Chen2020,Bandyopadhyay2020}.}
Such evolution 
{ corresponds to the observed} gradual decrease of the Alfv\'enic alignment between velocity and magnetic fluctuations \citep{bavassano1998,Bavassano2002}.
{ Alternatively, observations could result from the competing action between a coherent component (the intermittent inertial range structures generated by turbulence) with a stochastic component \citep[$1/f$-range propagating Alfv\'enic fluctuations,][]{Bruno2001,Bruno2003,Borovsky2008}. }
In both interpretations, the slow solar wind milder evolution is therefore associated with the reduced presence of Alfv\'enic fluctuations \citep{Tu1995,BrunoCarbone2013}.
{ Recent observations of Alfv\'enic solar wind closer to the Sun confirmed the radial evolution properties briefly described above \citep{Chen2020,Bourouaine2020,Hernandez2021}.}
The above observations of spectra and intermittency are generally used to support the evolving nature of the solar wind turbulence in the inner heliosphere, and to constrain global solar wind models and their energy budget. 


{ As an alternative to spectra, the turbulent cascade can be examined using the scaling properties of the third-order moments of the fluctuations \citep{Marino2023}. 
Based on robust theoretical predictions \citep[][]{Politano1998}, third-order laws allow to estimate the energy transfer rate of turbulence \citep{Sorriso-Valvo2007}. Under specific assumptions, this quantity represents a more fundamental measure of the state of turbulence, in comparison to power spectra and intermittency.} 
{ Furthermore, studies of radial evolution of turbulence mostly} rely on a statistical approach, so that different samples may include plasma from different originating coronal regions or solar activity, resulting in different initial energy injection, nonlinear coupling efficiency, or in-situ energy injection from instabilities or large-scale structures. 
These can all diversely contribute determining the properties and energetic content of the turbulent cascade.
One possible way to mitigate such inhomogeneity is to measure the same plasma with two spacecraft that are occasionally radially aligned at different distance from the sun~\citep[][]{DAmicis2010,Telloni2021-radial}. 
Alternatively, under optimal orbital configurations, plasma from a steady solar source can be measured by the same spacecraft at different times \citep[see, e.g.,][]{Bruno2003}. 

In this article, we study the status of the turbulence at different radial distances from the Sun using the third-order moment law and intermittency, as measured in a set of recurrent streams of solar wind measured by the Helios 2 spacecraft. { Based on a qualitative comparison with the plasma generated with a magnetohydrodynamic simulation in the spin-down phase, we interpret the observed trend of the energy transfer rate  in the solar wind as to be determined, among other process, by a decay of turbulence occurring with the heliocentric distance.}
In Sections~\ref{Sec:data} and~\ref{Sec:methods} we describe the dataset and the methodology used to investigate the status of turbulence. Section~\ref{Sec:results} provides the results of the analysis of Helios 2 data, while in Section~\ref{Sec:simula} results of a lattice-Boltzmann numerical simulations are shown. Finally, conclusions are given in Section~\ref{Sec:conclusions}.

\section{Helios 2 Data}
\label{Sec:data} 

\begin{figure*}
    \centering
    \includegraphics[width=0.9\textwidth]{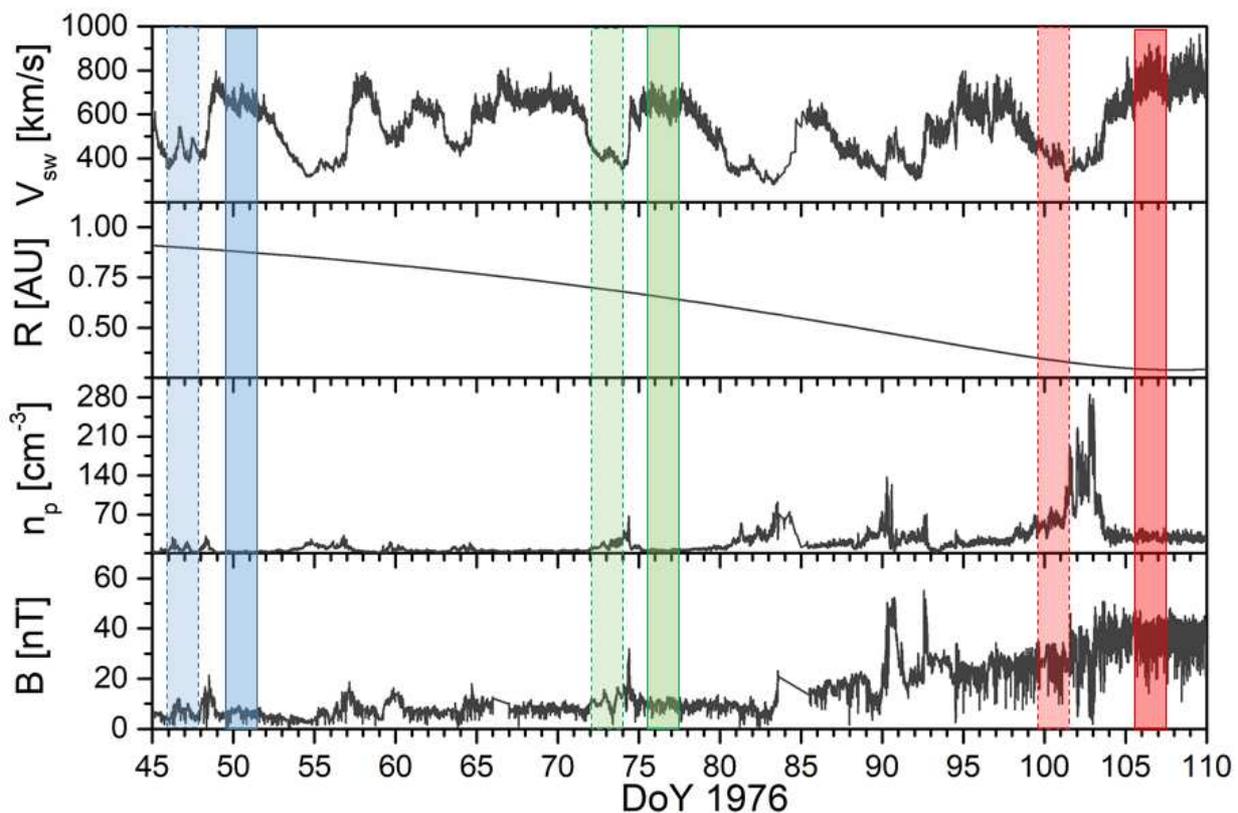}
    \caption{Helios 2 measurements during days 45---110 of 1976. From top to bottom: solar wind bulk speed $V_{sw}$, distance from the sun $R$, proton density $n_p$, and magnetic field magnitude $B$. Colored shaded areas identify the selected streams at 0.3 au (red), 0.7 au (green) and 0.9 au (blue). Lighter (darker) colors indicate slow (fast) streams.}
    \label{fig:data}
\end{figure*}

One popular case of measurements of solar wind from the same source occurred in 1976, when Helios 2 measured plasma and fields of three fast solar wind streams at different distances (0.9, 0.7 and 0.3 au) originated from a persistent polar coronal hole, which was reasonably stable during nearly two solar rotations~\citep[][]{bavassano1982b,Bruno2003}. 
Similarly, three preceding slow, non Alfv\'enic wind streams were used as samples of evolving slow solar wind.
These streams have provided outstanding information about the radial evolution of turbulence, since the initial conditions were statistically steady and no stream interactions were included, Note, however, that the more dynamical solar source of the slow streams might not be as steady as in the case of the coronal hole generating the fast wind. 
Figure~\ref{fig:data} shows an overview of solar wind bulk speed $V_{sw}$, spacecraft radial distance from the sun $R$, proton density $n_p$, and magnetic field magnitude $B$ during the days 45 to 110 of 1976. 
Six color-shaded areas identify the selected streams, at 0.3 au (red), 0.7 au (green) and 0.9 au (blue). For each distance, lighter colors and dashed lines indicate slow streams, darker colors and full lines identify fast streams.
%
%

\section{Methodology: the Politano-Pouquet law}
\label{Sec:methods}

Past studies of turbulence mainly relied on spectral and structure functions analysis~\citep[][]{bavassano1982b,bavassano1982a,Bruno1985,Bavassano2002,Bruno2003,Bruno2004,Bruno2004-levy,Bruno2014,Perrone2018}.
However, the nature of the turbulent cascade is more thoroughly captured by the scaling of the third-order moments of the fluctuations, an exact relation obtained from the incompressible MHD equations under the hypothesis of stationarity, isotropy, and large Reynolds' number \citep{Politano1998}. 
Called Politano-Pouquet (PP) law, it prescribes that the mixed third-order moment of the MHD fields fluctuations is a linear function of the scale:  
\begin{eqnarray}
    Y(\Delta t) &\equiv& \langle \Delta v_L (|\Delta \mathbf{v}|^2+|\Delta \mathbf{b}|^2) - 2 \Delta b_L (\Delta \mathbf{v}\cdot\Delta \mathbf{b})\rangle   \nonumber \\
    &=& \frac{4}{3} \varepsilon V_{sw} \Delta t \, .
\label{Eq:PP}
\end{eqnarray}
$\Delta \phi_L$ are longitudinal increments of the component of a generic scalar or vector component $\phi$ in the sampling direction, with the magnetic field $\mathbf{B}$ in velocity units through the mass density $\rho$, $\mathbf{b}=\mathbf{B}/(4\pi\rho)^{1/2}$.
The mean bulk speed $V_{sw}$ allows transforming spatial lags $\Delta l$ to temporal lags $\Delta t$ via the Taylor hypothesis, $\Delta l = -V_{sw} \Delta t$ \citep{Taylor1938}.
Observing linear scaling ensures that a turbulent cascade is active and fully developed, and that the complex hierarchy of structures on all scales is well formed and sustains the cross-scale energy transfer leading to small-scale dissipation. 
Measuring the energy transfer rate $\varepsilon$ provides a quantitative estimate of the turbulent energy flux. This is an invaluable information in the collisionless solar wind, where energy dissipation cannot be measured using viscous-resistive modeling. { The same quantity could in principle be obtained from the (Kolmogorov) spectrum, but its evaluation includes a constant factor that can hardly be obtained in the highly variable solar wind.} 
Third-order scaling laws \citep{Marino2023} have been used to determine the properties of turbulence in the solar wind at 1 au \citep{MacBride2005,Smith2009,Stawarz2010,Coburn2012}, in the outer \citep{Sorriso-Valvo2007,Marino2008,Carbone2009,Marino2012} and inner \citep{Gogoberidze2013,Bandyopadhyay2020,Hernandez2021,Wu2022} heliosphere, and in near-Earth \citep{Hadid2017,Quijia2021} and near-Mars \citep{Andres2020} environments. 
Data from Ulysses in the polar outer heliosphere \citep{Marino2008,Marino2012,Watson2022} and from Parker Solar Probe in the ecliptic inner heliosphere \citep{Bandyopadhyay2020} suggest that the energy transfer rate statistically decreases radially.
At the same time, the fraction of solar wind intervals where the linear scaling is observed increases { radially} \citep{Marino2012}, in agreement with the observed decrease of the cross-helicity and generally supporting the evolving nature of the turbulence in the expanding heliosphere.

\section{Results: mixed third-order moment scaling laws, turbulent energy transfer rate and intermittency}
\label{Sec:results} 

\begin{figure}
    \centering
    \includegraphics[width=0.48\textwidth]{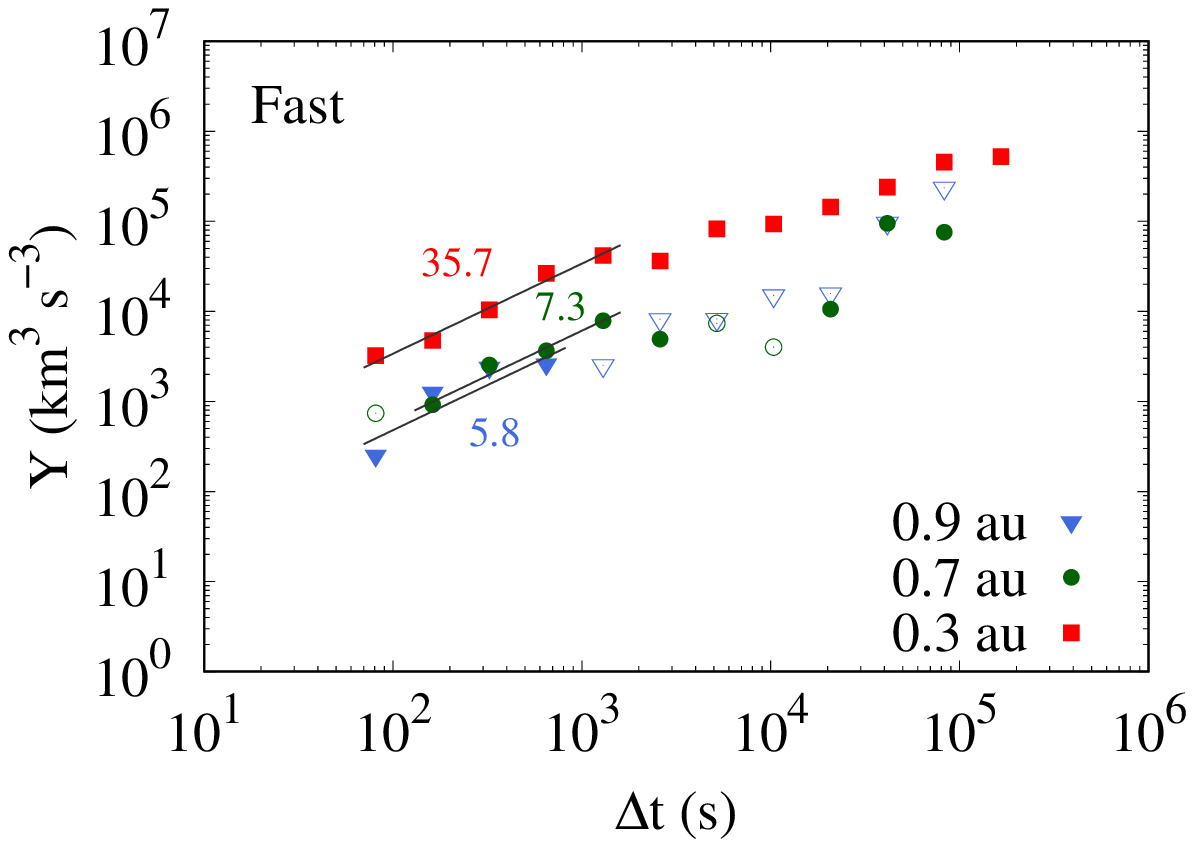}
    \includegraphics[width=0.48\textwidth]{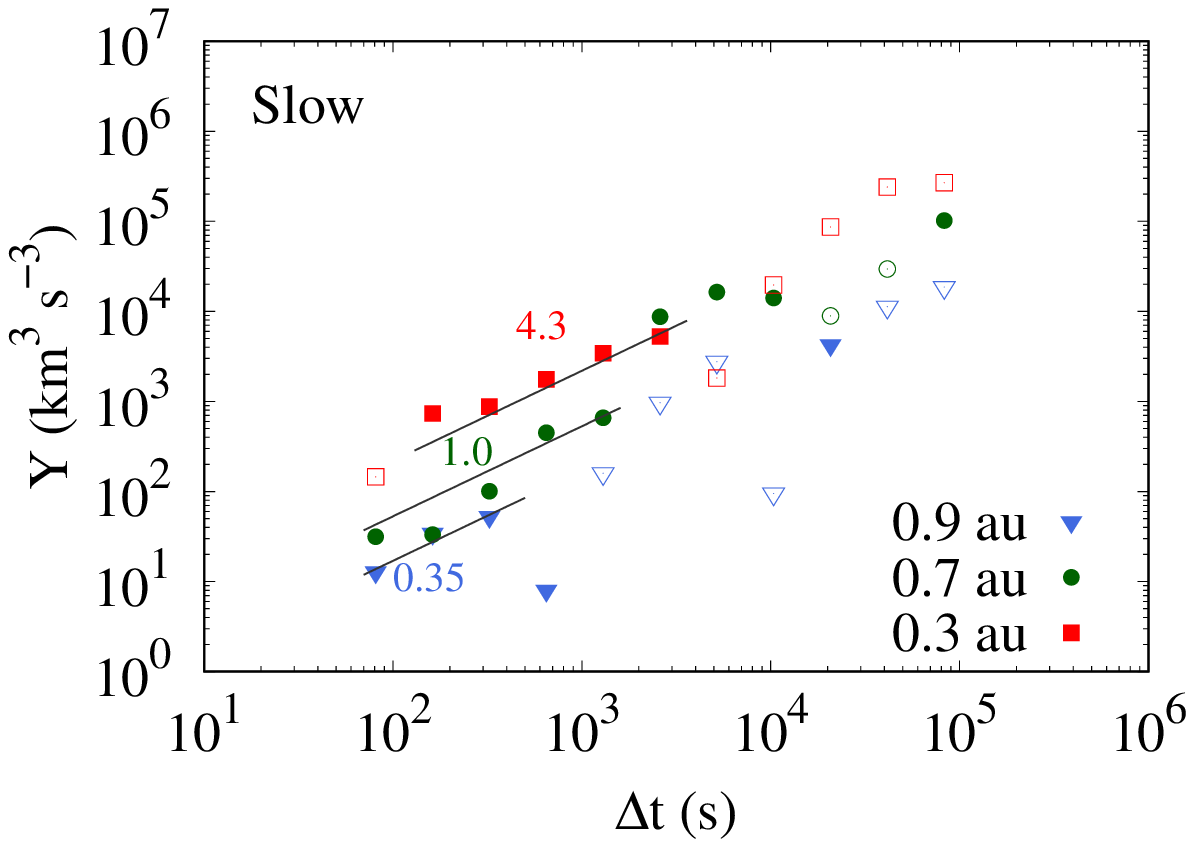}

    \caption{PP scaling for fast (top) and slow (bottom) streams, at three different distances from the sun.
    Linear fits are indicated (grey lines), along with the mean energy transfer rate, $\varepsilon$ (kJ kg$^{-1}$ s$^{-1}$) (color coded). Full and empty symbols refer to positive and negative values of $Y$, respectively.}
    \label{fig:Helios-yaglom}
\end{figure}

Here we present the analysis of the PP scaling law in the three fast and three slow streams described above, using the 81 s resolution plasma and magnetic field measured by Helios 2.
The mixed third-order moments, Equation~(\ref{Eq:PP}), are displayed in Figure~\ref{fig:Helios-yaglom} for each of the six intervals. Statistical convergence of the samples was tested using standard techniques~\citep[][]{Dudok2004,Kiyani2006}. 
For both fast and slow streams, the different colors indicate different distances, as stated in the legend. The mixed third-order moments $Y$ are mostly positive (full symbols), indicating direct energy transfer form large to small scales. Negative values, most probably due to lack of convergence or presence of large-scale velocity shears~\citep[][]{Stawarz2011}, are indicated by open symbols, and are not considered in this study. 
%
%
An inertial range was identified for each case, at timescales between 81 s and $\sim$20 minutes, although the linear scaling is better defined and more extended in the samples at 0.3 au.
The upper scales observed here are larger than, but roughly consistent with, the outer scale of the turbulence, estimated as the correlation timescale, $\tau_c$ \citep{Greco2012apj}, and partially include the $1/f$ range, as will be discussed below.
This shows that the six intervals can be considered as samples of fully developed turbulence.

Fitting the PP law provides the mean energy transfer rate, given in colors next to each fitted line in Figure~\ref{fig:Helios-yaglom}. 
The fit is performed on a range including more than one decade of scales in all cases except for the slow solar wind at 0.9 au, where a slightly shorter range is covered.
{ It should be observed that in the two examples studied here the linear scaling range is broader closer to the sun, while the scaling becomes less clear near 1 au.}

However, for the purpose of this study, the relevant information is the energy transfer rate, which is sufficiently well represented by the power-law fits shown in Figure~\ref{fig:Helios-yaglom}.
The first notable characteristic is that slow streams have smaller energy transfer rate than fast streams at all distances, suggesting that the initial energization of the turbulence is stronger in fast wind, perhaps not systematically but in the cases under study, or that its decay is faster in slow wind. 
{This is consistent with the known correlation between turbulence amplitude and both proton temperature and wind speed \citep[see, e.g.,][]{Grappin1991}.}
Furthermore, in both fast and slow wind the energy transfer rate consistently decreases with the distance, {as expected from the observed decay of the turbulent fluctuations \citep{BrunoCarbone2013}. }
Such observations seem to indicate that in the fast streams, while the spectrum broadens and the small-scale intermittent structures emerge \citep{bavassano1982b,Bruno2003}, the cascade transports less energy across the scales. 
The same decreasing energy transfer is observed in the otherwise steadier slow streams.
%
%
\begin{figure}
    \centering
    \includegraphics[width=0.48\textwidth]{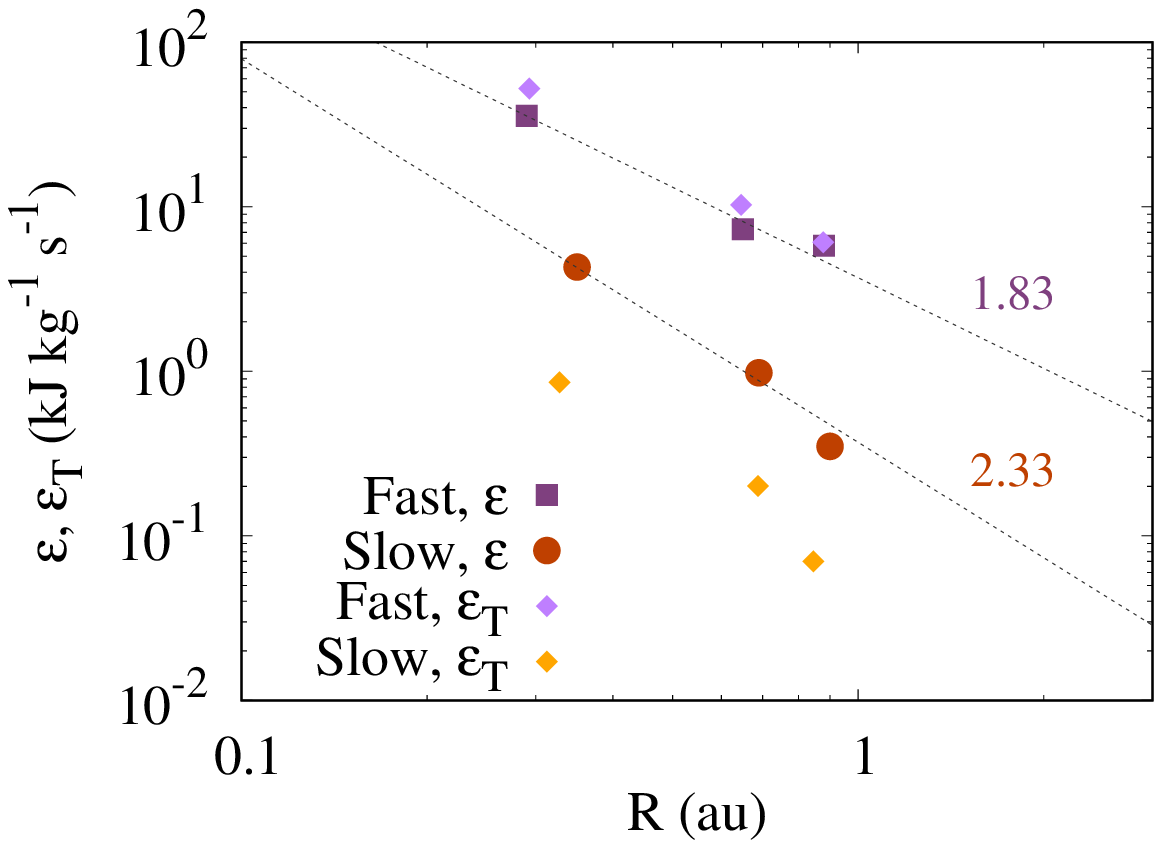}
    \includegraphics[width=0.48\textwidth]{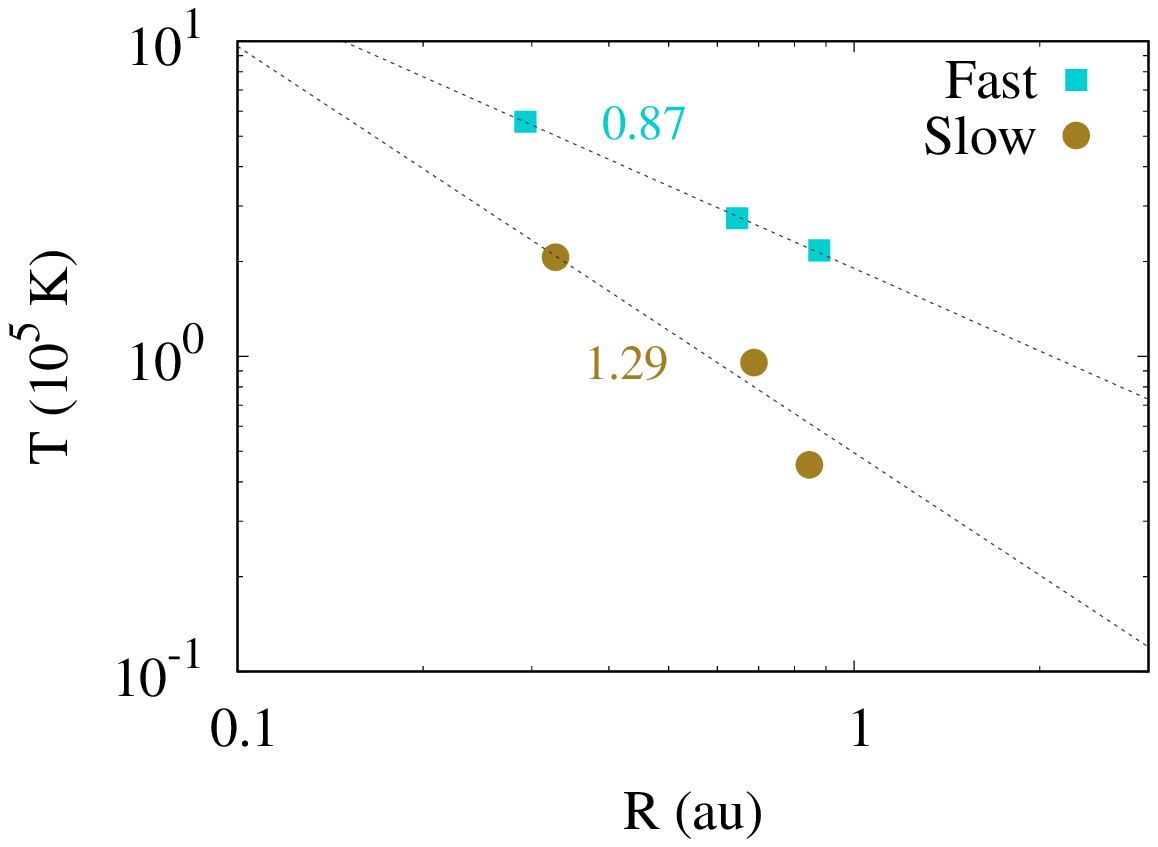} 
    \caption{Top panel: energy transfer rate, $\varepsilon$, versus the distance from the sun, $R$. 
    Fast streams are indicated with dark purple squares, slow streams with dark orange circles. Power-law fits and the relative scaling exponents are shown.
    The heating rate obtained using equation~(\ref{eq:vasquez}), $\varepsilon_T$, is also indicated with purple diamonds (fast streams) and orange diamonds (slow wind). 
    Bottom panel: the temperature decay for the fast (cyan squares) and slow (green circles) streams, with power-law fits and indicated exponents.}
    \label{fig:Helios-epsilon}
\end{figure}
In the top panel of Figure~\ref{fig:Helios-epsilon}, the energy transfer rate is plotted versus the distance from the Sun.
The measured values can be compared with standard estimates of the proton heating rate \citep[][]{Marino2008,Marino2011}. This can be obtained through a simple model that, under the isotropic fluid approximation, neglecting heat fluxes, and assuming a stationary flow, uses the solar wind bulk speed and temperature radial  profiles to determine the proton heating rate accounting for deviation from adiabatic cooling~\citep[][]{Verma2004,Vasquez2007}. 
In particular, if the proton temperature decays as a power law of the distance, $T(R)\sim R^{-\xi}$, then the proton heating rate is given by
 \begin{equation}
 \varepsilon_T = \frac{3}{2}\left( \frac{4}{3}-\xi \right)\frac{k_B V_{SW}(R) T(R)}{R m_p} \, ,
 \label{eq:vasquez}
 \end{equation}
where $k_B$ is the Boltzmann constant and $m_p$ is the proton mass. The decay exponent $\xi$ of the temperature can be estimated using Helios 2 data. 
For the two sets of fast and slow streams, the mean temperature for each stream is shown in the bottom panel of Figure~\ref{fig:Helios-epsilon}. 
Assuming power-law decay, a fitting procedure provides $\xi = 0.87\pm 0.01$ for the fast streams, and $\xi = 1.29 \pm 0.03$ for the slow streams. 
These considerably deviate from the expected adiabatic value ($\xi = 4/3$) for the fast wind, while for the slow streams the cooling is closer to adiabatic. 
{ \citet{Totten1995} used Helios 1 data to obtain radial profiles of the proton temperature, which was observed to consistently decay with exponent $\xi\simeq 
0.9$, independent of the wind speed. 
{Note that the results shown here might be specific to the case under study, while in \citet{Totten1995} a large statistical sample was used.}
Similar results were found using MHD numerical simulations \citep{Montagud-Camps2018,Montagud-Camps2020}. The Helios 2 results shown here are in good agreement with the above observations and models for the fast wind streams, but not for the slow ones.}

Using the obtained temperature radial decay exponent and the mean speed and temperature of the three streams, the radial profile of the approximate heating rate can be estimated. The values obtained are shown in the top panel of Figure~\ref{fig:Helios-epsilon}. For the fast streams, the agreement with the estimated energy transfer rate is excellent. This demonstrates that the energy transfer rate estimated using the incompressible, isotropic version of the PP law are sufficiently accurate. On the other hand, for the slow wind the required heating is nearly one order of magnitude smaller than the observed turbulent heating rate. This is a consequence of the very weak deviation from adiabatic of the power-law temperature decay. Possible reasons for this discrepancy include: (i) the relatively poor power-law profile of $T(R)$ and subsequent underestimation of the required heating rate; (ii) the possible variability in the slow wind source region, which would affect the stationarity assumption for the model and result in unaccounted for temperature variability; (iii) energy lost to heat flux and electron heating, not included in equation~(\ref{eq:vasquez}), and which might be more relevant than for the Alfv\'enic, fast wind. Nevertheless, even for the slow streams  the decay of the predicted heating rate is close to the observed decay of the turbulent energy transfer rate.

In fact, for plasma proceeding from an approximately stationary coronal structure and far from stream interaction regions, the turbulent energy transfer can be expected to decrease as a power law of the distance. 
If we assume for the energy transfer rate a power-law radial decay, $\varepsilon \sim R^{-\alpha}$, the measured values can be fitted to power-laws, providing the decay exponents $\alpha_F\simeq1.8\pm 0.2$ and $\alpha_S\simeq2.3\pm 0.2$ for fast and slow streams, respectively. 
The slower decay observed in the fast streams could be the result of the local energy injection from the $1/f$ reservoir, which may partially compensate the dissipation losses and thus slightly slow down the decay in comparison with the slow wind, where no supplementary injection is provided. 

\begin{figure}
    \centering
    \includegraphics[width=0.48\textwidth]{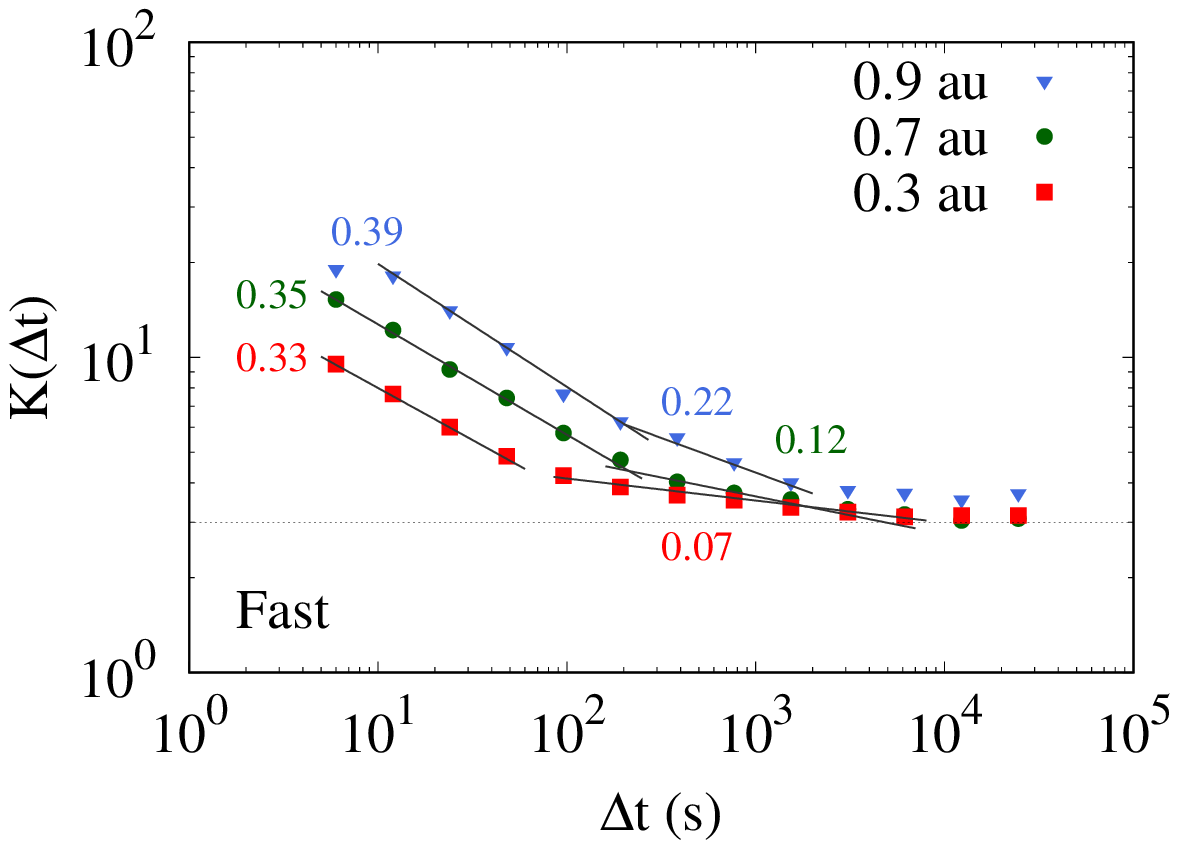}  
    \includegraphics[width=0.48\textwidth]{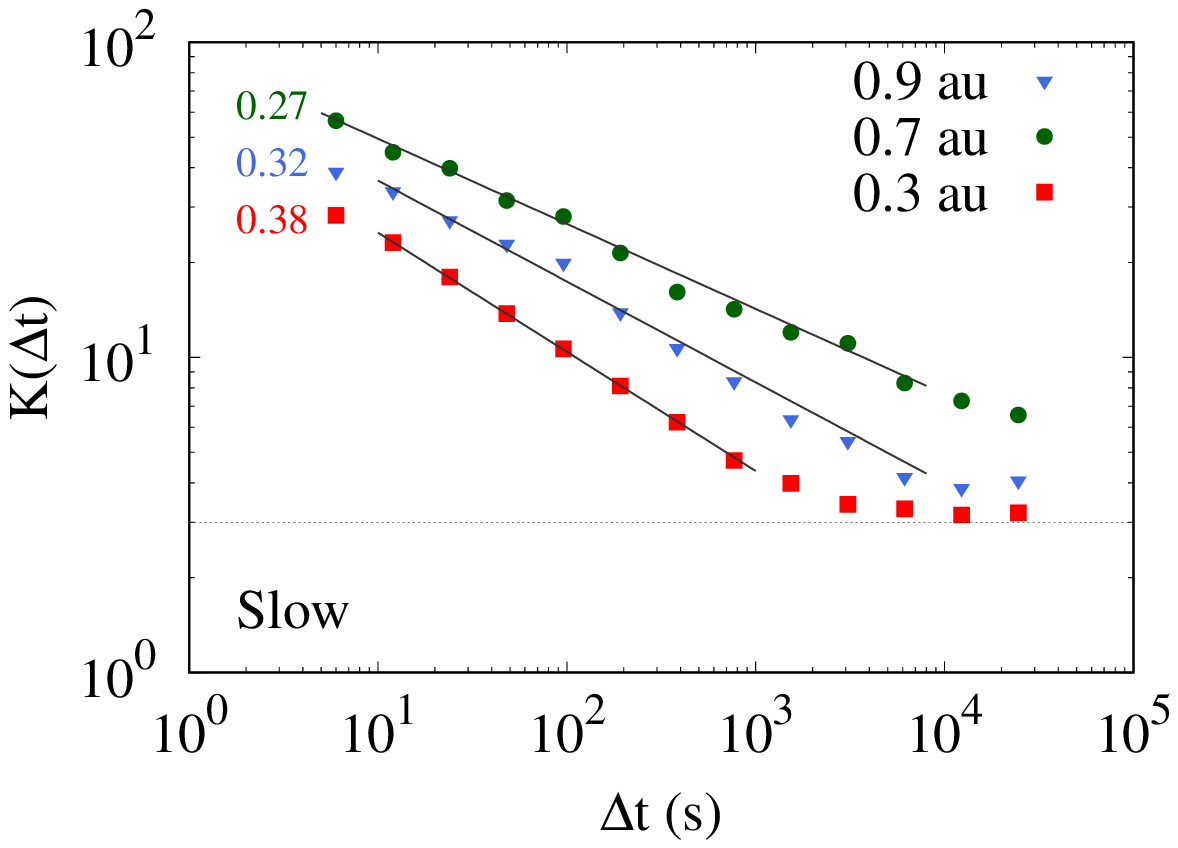}  
    \caption{Magnetic field kurtosis, $K$, for fast streams (top) and slow streams (bottom), averaged over the field components. 
    Power-law fits in the 5/3 and $1/f$ spectral ranges are indicated with the fitted scaling exponents (fitting errors are $\leq0.01$).}
    \label{fig:kurt}
\end{figure}

{ The above scenario provides a quantitative estimate of the turbulence decay observed in undisturbed expanding samples, such as the Helios 2 recurrent streams. 
Such observation could be useful to constrain models of the radial evolution of turbulence. These may include, for example, the simple damping of both velocity and magnetic fluctuations due to the conservation of angular momentum and of magnetic flux in an expanding plasma volume advected by the radial wind \citep{Parker1965,Heinemann1980}, as well as local energy injection as it results from expansion and large-scale shears \citep{Velli1990,Grappin1993,Tenerani2017}, or modification of the timescale of nonlinear interactions associated with the radial decrease of Alfv{\'e}nicity \citep{Smith2009,Stawarz2010}.}
{Since the results presented here are based on only two case-study events, they lack  generality. On the other hand, they do provide more rigorous parameters for the specific intervals. A larger study of similar events, studied  individually, is however necessary to cover a broader range of parameters.}
%
%
In order to complete the turbulence analysis of the streams under study, we quantify the degree of intermittency of the turbulent fluctuations. 
{ Intermittency is typically described by the scaling of the statistics of the fields fluctuations, measured through probability distribution functions \citep{Sorriso-Valvo1999,Pagel2003}, structure functions \citep{Carbone1994,Kiyani2009pre}, or multifractal analysis \citep{Macek2012,Alberti2019}.
A standard estimator of intermittency is provided by the power-law scaling of the kurtosis} of a magnetic field component $B_i$, $K(\Delta t)=\langle\Delta B_i^4\rangle/\langle\Delta B_i^2\rangle^2\sim \Delta t^{-\kappa}$.
The scaling exponent, $\kappa$, provides a quantitative measure of intermittency, being related to the fractal co-dimension of the intermittent structures \citep{Castaing1990,Carbone2014,Sorriso-Valvo2015}.
For these intervals, the kurtosis was already presented by \citet[][]{Bruno2003}. Nevertheless, here we perform a more detailed, quantitative study that will provide additional description of the intermittency.
The two panels of Figure~\ref{fig:kurt} show the magnetic field kurtosis for the three fast (top panel) and slow (bottom panel) solar wind streams, computed using 6-second cadence magnetic vectors and averaging over the three field components (after verifying that all individual components displayed similar behaviour).
In the fast streams, two power-law scaling ranges are identified in the inertial range, $\sim$6---200 s, { and at lower frequency, $\sim$200---8000 s. 
In agreement with spectral observations, the break between the two ranges migrates towards larger scales with increasing distance from the Sun~\citep[see, e.g.,][and references therein]{BrunoCarbone2013}. 
Note that, for the streams at 0.3 and 0.7 au, the low-frequency range mostly includes the $1/f$ spectral range.}
Power-law fits provide the scaling exponents $\kappa$, which are indicated in the figure.
The inertial-range exponents agree with previous observations \citep{DiMare2019,Sorriso-Valvo2021}, while the smaller $1/f$-range values are closer to fluid turbulence's~\citep[where exponents are typically around 0.1; see, e.g.,][]{Anselmet1984}.
In both ranges, the values of $K$ and the scaling exponents quantitatively confirm the radial increase of intermittency \citep{Bruno2003}. 
{ To the best of our knowledge, the kurtosis' power-law scaling in the $1/f$ range was not observed before. 
Jointly with the observation of the PP law discussed above, it suggests that, at least in the streams closer to the Sun, nonlinear interactions are effectively transferring energy across scales, even in the $1/f$ range}. 
This observation opens interesting questions about the nature of the fluctuations in the low-frequency range, which calls for more detailed studies.
{ On the other hand, in the slow wind streams, where no $1/f$ spectral range is observed, a single power-law covers the whole range}, with no clear radial trend. The scaling exponents reveal strong intermittency at all distances.
%
%

\section{Comparison with numerical simulations of unforced MHD turbulence}
\label{Sec:simula} 

\begin{figure*}
    \centering
    \includegraphics[width=0.95\textwidth]{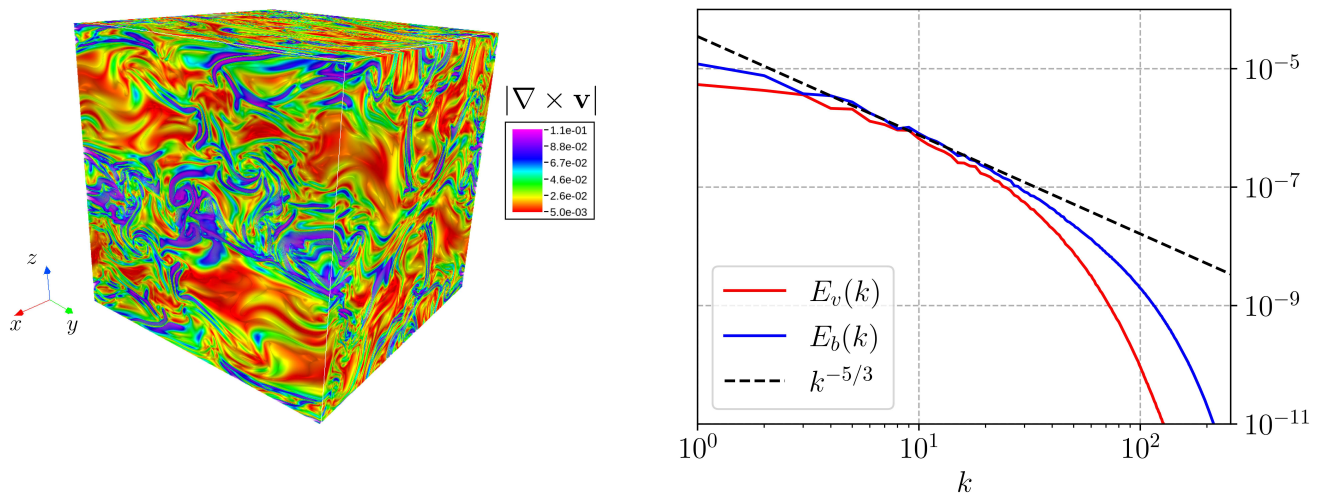}
    \caption{Left panel: rendering of the vorticity field, at the time $t^*$ when the peak of the density current is reached.
    Right panel: isotropic kinetic and magnetic energy spectra for run I in the table I of the main text at $t^*$. 
    }
    \label{fig:spectra}
\end{figure*}

The power-law radial decay of the turbulent energy transfer rate and the associated increase of intermittency, highlighted in the previous section, represent a solid constraint for models of solar wind turbulence. 
It is interesting to notice that similar properties are also typical features of unforced fluid and magneto-fluid turbulence, as observed in numerical simulations \citep{Biskamp1993,Hossain1995,Miura2019,Bandyopadhyay2019decay}. 
Using a decaying direct numerical simulation integrating the incompressible MHD equations, with no mean magnetic field \citep{Foldes2022}, here we show how a simplified framework not incorporating signature features of solar wind turbulence (for example anisotropy), not accounting for effects of the solar wind expansion \citep[][]{Grappin1993,Verdini2015,Tenerani2017}, 
{ is able to qualitatively reproduce trends and statistics observed by Helios, thus can be used to decipher the phenomenology underlying the turbulent energy transfer in the solar wind. What emerged from our analysis is that the phenomenology described in the previous sections is compatible with the temporal decay of a magnetohydrodynamic unforced plasmas. 
In other words, the basic three-dimensional MHD simulations used here provide indications of qualitative similarity between the radial evolution of turbulence in the expanding solar wind and the temporal decay of unforced MHD turbulence via viscous-resistive dissipation.}
%
%
\begin{table*}
\centering
\begin{tabular}{c|cccccc}
\hline
run  &  $\mathrm{Re=Re_m}$ &  $v_{rms}^*$        & ${v_{rms}^0}/{b_{rms}^0}$ & $\mathrm{Ma}$ & $t^*$ & $\xi$  \\ \hline
I    &  $\sim$ 1500            & $8.5\cdot 10^{-4}$  &      $\sim 0.5$           &   0.005       &  79.5 &   2.4      \\
II   &  $\sim$ 900             & $2.7\cdot 10^{-3}$  &      $\sim 1.0$           &   0.005       &  51.9 &   1.7      \\
III  &  $\sim$ 800             & $2.5\cdot 10^{-3}$  &      $\sim 2.0$           &   0.001       &  45.3 &   1.8      \\ \hline
\hline
\end{tabular}
\vskip 12 pt
\caption{Adimensional parameters at the current peak  $t^*$ (in turnover time units). $\mathrm{Re}$ and $\mathrm{Re_m}$ are respectively the Reynolds and magnetic Reynolds numbers; $\mathrm{Ma}$ is the Mach number; $v_{rms}^0/b_{rms}^0$ is the ratio between the $rms$ velocity and magnetic fluctuations at the initial time of the simulation ($t=0$);
$\xi$ is the fitted exponent (see Figure~\ref{fig:LB-epsilon}). 
Box size, $(2\pi)^3$, and magnetic Prandtl number, $Pr_m=1$, are the same for all runs.}
\label{tab:run}
\end{table*}
{ This is not in contrast with the expanding-box simulations previously suggesting that the inclusion of expansion effects results in faster energy decay and in inverted cross helicity radial profile, that switches from increasing to decreasing with distance \citep[see, e.g.,][]{Dong2014,Montagud-Camps2020,Montagud-Camps2022}.
In this work, we use actually a lattice Boltzmann (LB) code, FLAME \citep{Foldes2022}, to integrate the quasi-incompressible MHD equations in a three-dimensional periodic domain, not expanding. In the LB approach, the volume discretization is operated on a gas of particles distributed on a lattice, rather than on a grid. 
The dynamics of the particles develops in the frame of the kinetic theory, the temporal evolution of the plasma being achieved through the recursive application of simple collision and streaming operations. 
The macroscopic MHD fields (e.g. fluid velocity $\mathbf{v}$, density $\rho$, and magnetic field $\mathbf{B}$) are then obtained through integration of the statistical moments of the particle distribution functions. Please note that the particles here are not plasma particles, they exist at the level of the numerical scheme and are instrumental to the LB approach to obtain (in the case of FLAME) the fields of the simulated magnetohydrodynamic plasma \citep[see details in][]{Foldes2022}.}
We examine three runs, whose parameters are listed in Table~\ref{tab:run}. The latter are initialized with a standard Orszag-Tang (OT) vortex \citep{OT,Mininni2006}, using different kinetic-to-magnetic energy ratio, $v_{rms}^0/B_{rms}^0$, where $v_{rms}^0$ and $B_{rms}^0$ are the initial rms values of the field fluctuations \citep[for the Helios 2 solar wind intervals studied here, $v_{rms}/B_{rms}\sim$ 0.5--1,][]{Bruno1985}. 
The integration is performed over a $512^3$-point three-dimensional lattice \citep{OT,Mininni2006}, the functional form of OT being:
$\mathbf{U}(\mathbf{x},0) = v_{rms}^0 \left[-2sin(y),2sin(x),0\right]$ and $\mathbf{B}(\mathbf{x},0) = B_{rms}^0 \left[-2sin(2y)+sin(z),2sin(x)+sin(z),sin(x)+sin(y)\right]$.
Before computing any statistics, the simulation is let to evolve until the plasma reaches a state of fully developed turbulence when the volume averaged density current ,$\langle j\rangle$, attains its peak value, at the time $t^*$. 
A snapshot of the vorticity field at $t^*$ is shown for run I in Figure \ref{fig:spectra}, along with the corresponding isotropic magnetic and kinetic spectral trace. %
{ For all runs, spectra exhibit the typical extended power-law inertial ranges, which enables qualitative comparison to the SW.}

\begin{figure}
    \centering
    \includegraphics[width=0.48\textwidth]{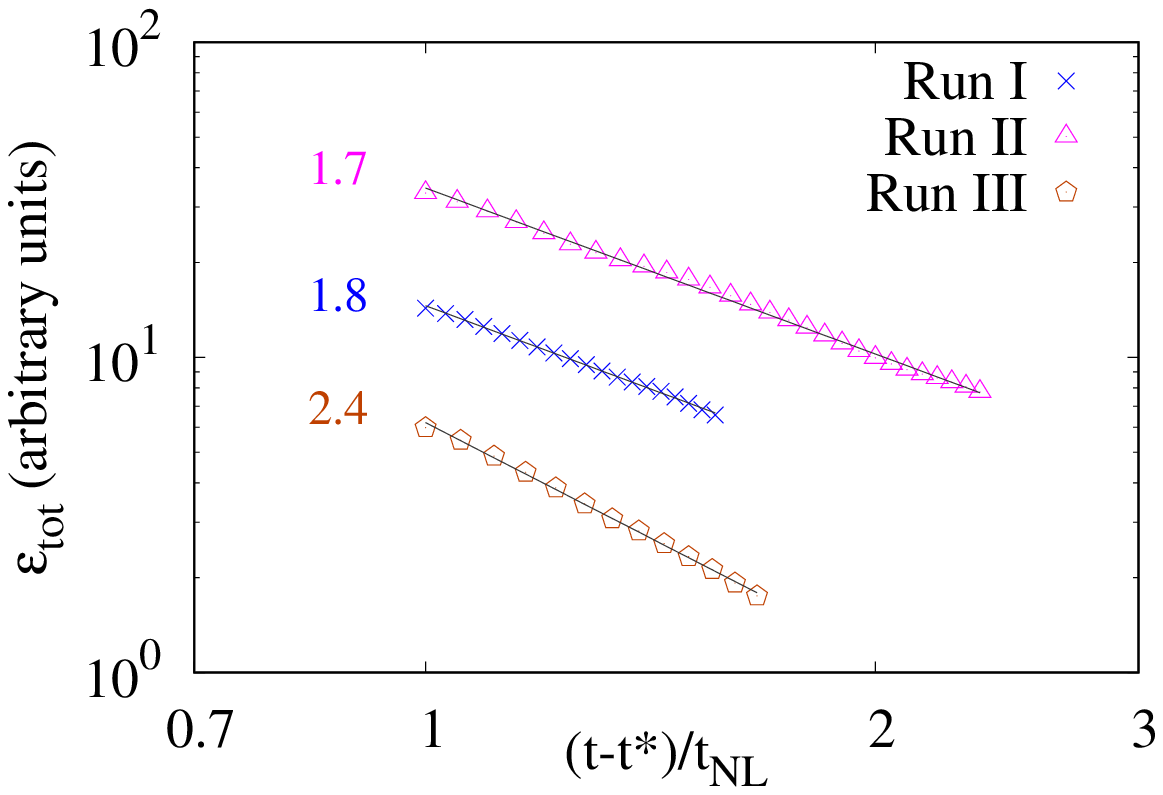} 
    \includegraphics[width=0.48\textwidth]{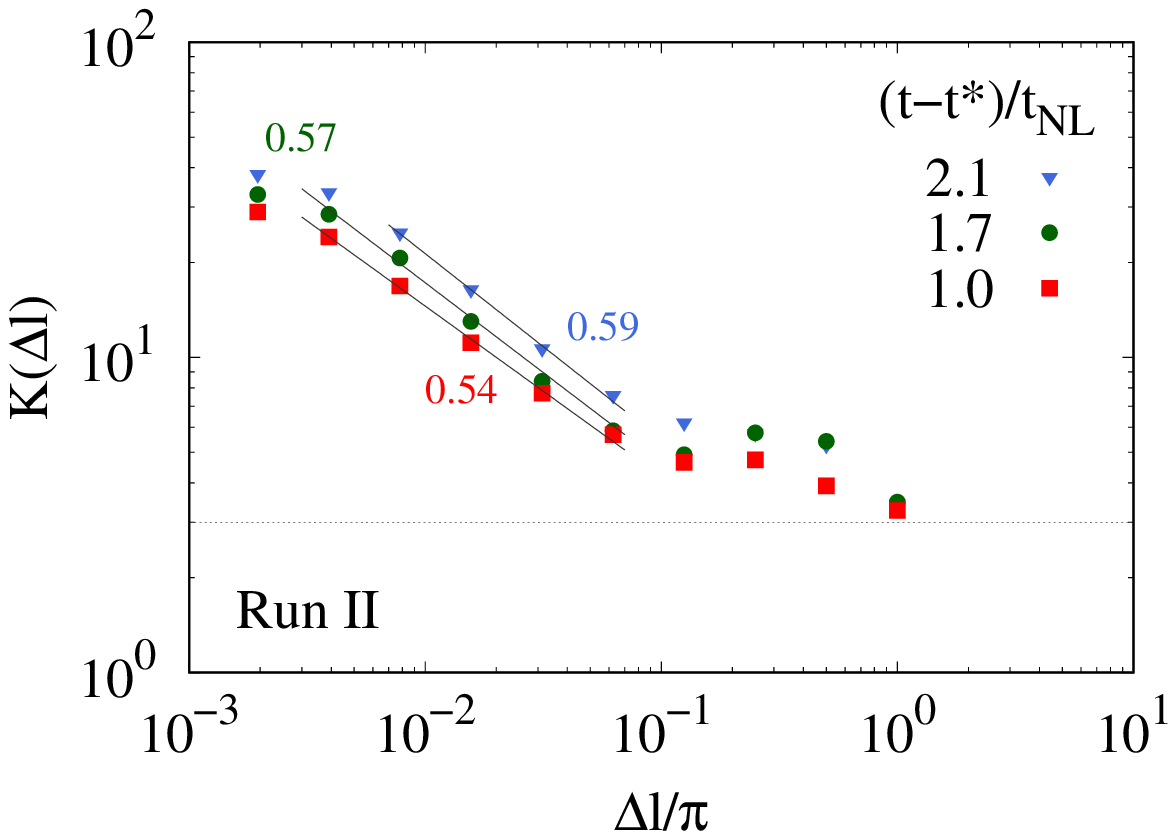}
    \caption{Top panel: energy dissipation rate, $\varepsilon_{tot}$ (arbitrary units), versus the rescaled simulation time, $(t-t^*)/t_{NL}$, for the three simulation runs.
    Bottom panel: magnetic field kurtosis, $K$, versus the spatial increment $\Delta l$, averaged over the field components, shown here for run II at three different times (see legend). 
    For all plots, power-law fits and the relative fitted exponents are indicated (fitting errors $\leq0.01$).}
    \label{fig:LB-epsilon}
\end{figure}

{ It must be pointed out that, since the solar wind speed is reasonably steady, the radial distance of each of the six Helios 2 streams analyzed here can be ideally converted to time of travel from an arbitrary initial position where the turbulence peaks, $R_0$. We can expect this to be close to the Sun \citep{Bandyopadhyay2020}, for example at the Alfv\'en point \citep{Kasper2021,Zhao2022}, or it can be identified with the stream under study closest to the Sun ($R_0=0.3$ au). Further expressing the solar wind time of travel in units of the initial nonlinear time, $t_{NL}$ (here taken as the average between characteristic kinetic and magnetic nonlinear times, respectively $t_{v}=L_{int}^v/v_{rms}^*$ and $t_{B}=L_{int}^B/B_{rms}^*$, where $L_{int}^v$ and $L_{int}^B$ are the kinetic and magnetic integral scales, whereas $v_{rms}^*$ and $B_{rms}^*$ are the rms values of the field fluctuations, all estimated at $R_0$) enables the comparison between observational and numerical estimates and statistics. 
This suggests the possible use of the following simple expression to determine the normalized ``age'' of the turbulence: $t_{turb} = ( R - R_0 )/( V_{sw} t_{NL} )$.
We will not make use of this parameter since our comparison with the numerical simulation results is only qualitative. Furthermore, determining $R_0$ and the integral scales is not trivial in solar wind data. However, since the parameters in the above transformation (the solar wind speed, the initial distance, and the nonlinear time at the initial position) can all be considered as constant, the power-law scaling presented will not be affected by this transformation.}
{ Computing spectral energy fluxes from numerical simulation requires the integration of quantities over extended intervals, during which the system is assumed to be stationary. This condition can hardly be attained when dealing with spin-down runs. Moreover we would like to monitor the evolution of the turbulent energy transfer rate, in all the simulations in Table~\ref{tab:run}, at different times after the plasma has reached the density current peak. For this reason, here we make the assumption (reasonable in case of fully developed turbulence) that the rate at which kinetic and magnetic energies are transferred throughout the inertial range, in subsequent time intervals, as the system relaxes due to viscous effects, does follow the same trend of the small-scale dissipation with the evolutionary time of the simulations.}
{ We have thus computed systematically the volume-averaged total dissipation rate, $\varepsilon_{tot}=\varepsilon_V+\varepsilon_B$, where $\varepsilon_V=\nu\langle({\nabla v})^2\rangle$ and $\varepsilon_B=\eta\langle j^2\rangle$ are the kinetic and magnetic dissipation, respectively, and $\nu=\eta$ are the kinematic viscosity and the resistivity. 
Figure~\ref{fig:LB-epsilon} (top panel) displays the temporal evolution of the dissipation rate $\varepsilon_{tot}$ in runs I-III, starting from the turbulence peak $t^*$.  
All times are expressed in units of the nonlinear time, $t_{NL}$, estimated as described above using the simulation parameter computed at the time of the peak of the density current, $t^*$. 
For all runs, a power-law time evolution of the energy transfer rate can be clearly identified \citep{Batchelor1948,Hossain1995}, with fitted scaling exponents compatible with those observed in the solar wind. }

%

{ Finally, the bottom panel of Figure~\ref{fig:LB-epsilon} shows the magnetic field kurtosis versus the scale $\Delta l$ for Run II, at three different times $t>t^*$ in the simulation. The power-law scaling exponents indicate increasing intermittency with time \citep{Biferale2003}, similarly to observations in the fast recurrent stream studied here. The same trend is observed for all runs.}

{ By qualitatively comparing both the energy transfer rate and intermittency, evident similarities arise between the time evolution in our simulations of decaying MHD turbulence and the radial profile in the solar wind streams.
Although the numerical model used here is not intended to fully reproduce the solar wind features, such similarities suggest that the ongoing dissipation of turbulent fluctuations could concur to determine the observed radial evolution of solar wind turbulence.}

\section{Conclusions}
\label{Sec:conclusions}

We used Helios 2 measurements of the solar wind emitted from a steady coronal source and without interactions with coronal or heliospheric structures, collected at different distances from the Sun. 
We have shown that the turbulence energy transfer rate decays approximately as a power law of the distance, { and we provided measured decay exponents that may be used to constrain models of solar wind expansion.} 
{ It should be pointed out that the linear scaling range becomes narrow at larger distance from the Sun. In the slow wind, for example, such range covers slightly less than a decade, which is in contrast with the observed broad spectral inertial range. This is likely due to the limited statistics provided by the relatively low resolution Helios data, and more in general to the difficult observation of signed third-order scaling laws. Possible other reasons include the violation of the isotropy assumption and the presence of large-scale inhomogeneities \citep{Stawarz2011,Verdini2015anisotropy}.}
In the Alfv\'enic fast wind, the turbulence decay is also associated with increasing intermittency.  
The observations presented here are qualitatively compared with three-dimensional direct numerical simulations of decaying MHD turbulence. Despite the simulations used here do not include important elements such as the radial expansion of the solar wind, the observed similarity between trends of energy transfer and dissipation rates (estimated from Helios 2 observations and DNS, respectively) supports the possible relevance of dissipation in the radial evolution of solar wind turbulence.
Furthermore, the observation in the $1/f$ range of both the PP law and power-law scaling of the kurtosis suggests that, in fast solar wind, a turbulent cascade is active also at large scales, even in the presence of strongly Alfv\'enic large-scale fluctuations \citep{Verdini2012}. 
The behavior highlighted by our analysis, 
together with the observed parameters, can be relevant to constrain models of turbulence in the expanding solar wind and of the plasma heating observed in both fast and slow streams \citep{Marino2008,Marino2011}. 
Coordinated studies of PSP and Solar Orbiter measurements will add statistical significance to our observations \citep{Velli2020,Telloni2021}.

\begin{acknowledgements}
L.S.-V. and E.Y. were supported by SNSA grants 86/20 and 145/18. L.S.-V. is supported by the Swedish Research Council (VR) Research Grant N. 2022-03352. R.M. and R.F. acknowledge support from the project ``EVENTFUL'' (ANR-20-CE30-0011), funded by the French ``Agence Nationale de la Recherche'' - ANR through the program AAPG-2020. This research was supported by the International Space Science Institute (ISSI) in Bern, through ISSI International Team project \#556 (Cross-scale energy transfer in space plasmas).
\end{acknowledgements}

\bibliographystyle{aa}
\bibliography{AA-HELIOS}

\end{document}